\documentclass[epjCONF,columns]{svjour} 
\usepackage{lineno}
\usepackage{graphics}
\usepackage{subfig}
\usepackage[varg]{txfonts} 
\usepackage[latin1]{inputenc}
\session-title{2011 Hadron Collider Physics symposium}
\begin{document}
%
\title{A search for $t\bar{t}$ resonances in the dilepton channel in 1.04 fb$^{-1}$ of pp collisions at $\sqrt{s} = 7$ TeV with the ATLAS detector}
\author{S. Swedish \inst{1,2}\fnmsep\thanks{\email{swedish@cern.ch}} on behalf of the ATLAS Collaboration.}
\institute{Triumf, BC, V6T 2A3, Canada  \and University of British Columbia, BC, V6T 1Z4, Canada}
\abstract{
The first ATLAS result on a search for a high mass top pair resonance at the LHC, in the subset of
 events where both W bosons from the top decays decay to either a final state electron or muon, is presented.  The 
analysis is performed on 1.04$fb^{-1}$ of pp collisions at $\sqrt{s} = 7$ TeV.   Numerous models predict the 
production of new massive particles that decay preferentially to a top-anti-top pair, including Randall-Sundrum models 
where the observation of a Kaluza-Klein excitation of the gluon may be the first indication of the existence of an extra spatial dimension.
 In the analysis presented, a simple observable, sensitive to resonance mass, is formed by 
summing the missing transverse energy, and the transverse momenta of the selected jets and the two candidate 
leptons.  A deviation from the Standard Model prediction for this observable is searched for using Bayesian 
statistical methods that compare the yields, and shapes, of the Standard Model background and signal 
predictions for KK-gluons with masses between 500 and 1600 GeV.  No excess over the Standard Model is 
observed and 95\% C.L. upper limits are set on the production cross-section times branching 
ratio to top quarks for KK-gluon-like resonances.   The results of the analysis exclude Randall-Sundrum 
KK-gluons with masses less than 840 GeV. 
} 
\maketitle
\section{Introduction}
\label{intro}
The top quark holds a peculiar position in the Standard Model as the most massive known particle and the only fermion with mass 
close to the electroweak symmetry breaking scale.  As a result, the top quark has a special role in many scenarios that go beyond the 
Standard Model (SM), many of which predict new particles that decay preferentially to top anti-top quark pairs \cite{refTopSpec}.  One such scenario is 
the Randall-Sundrum model with a single finite extra-dimension (RS1) where the SM fermions and gauge bosons can propagate in the bulk.  In this scenario the observation 
of the first Kaluza-Klein (KK) excitation of the gluon may be the first indication of the existence of an extra spatial dimension \cite{ref1}.  
Such a new particle would manifest itself at the LHC as a massive $t\bar{t}$ resonance. 

This note describes an analysis that searches for a KK-gluon-like $t\bar{t}$ resonance at the LHC with the ATLAS experiment\footnote{For a more detailed description of this analysis see Ref. \cite{ref2}.} \cite{ref2,ref:atlasexp}.
We study the case where the KK-gluon decays into top anti-top pairs that further decay to leptons (electron, muon, or tau), and look in final states with 
reconstructed electrons and muons. The presence of two neutrinos from the $W$ decays in the final state complicate full mass reconstruction.
The scalar sum of the transverse momenta, $p_T$, of selected leptons and jets, denoted $H_T$, is measured and combined with 
the missing transverse energy expected from the 
neutrinos to form the variable $H_{T} + E_{T}^{miss}$.

$H_{T} + E_{T}^{miss}$ is correlated to the invariant mass of the event and provides some discrimination between resonance and SM $t\bar{t}$ events.  A deviation from the SM expectation in the $H_{T} + E_{T}^{miss}$ 
spectrum is searched for using a template shape fitting method for a range of resonance masses.  In the absence of signal, a
limit is set on the production cross-section times branching ratio, $\sigma B$, and this limit is converted into a mass limit on the KK-gluon in the RS1 model.

\section{Event Selection}
\label{EventSelection}
This analysis is conducted on data corresponding to an integrated luminosity of 1.04 fb$^{-1}$ that was collected from March to July 2011.  The event 
selection relies on reconstructed leptons, jets, and $E_{T}^{miss}$.
Lepton candidates are required to be isolated in order to suppress backgrounds from hadron and heavy flavour decays inside jets.  Electron 
candidates are required to have $E_T > 25$ GeV and $|\eta| < 2.47$.  In order to avoid the barrel-endcap transition region in the ATLAS calorimeter, 
electrons with $1.37 < |\eta| < 2.47$ are excluded.  Muon candidates are required to have transverse momentum $p_T > 25$ GeV, and $|\eta| < 2.5$.  
Candidate jets must have $p_T > 25$ GeV and $|\eta| < 2.5$.  

The event selection is motivated by the final state of the top anti-top decays, where events will contain two jets, two oppositely charged leptons, and at least two neutrinos.  Events are first preselected by requiring they meet the following criteria:

\begin{itemize}
\item Satisfy good data quality requirements. 
\item Pass either single electron or muon triggers with thresholds $E_{T} > 20$ GeV 
and $E_{T} > 18$ GeV respectively.  
\item Contain exactly 2 oppositely charged leptons (electrons or muons).
\item Have dilepton mass, $m_{ll} > 10 GeV$.
\item Contain two or more jets.  
\end{itemize}

Within the pool of preselected events, control and signal regions are defined.  For the control region a sample depleted of signal 
and dominated by Z+jets events is defined by placing an additional cut on the invariant mass of the two leptons, 
$|m_{\ell\ell} - m_{Z}|< 10$ GeV, and is restricted to the $ee$ and $\mu\mu$ channels. 

The definition of the signal region depends 
on the dilepton channel. The $ee$ and $\mu\mu$ channels require a cut of $E_{T}^{miss} > 40$ GeV and exclude the control region,
 $|m_{\ell\ell} - m_{Z}| >10 GeV$.  The $e\mu$ channel requires  
$H_{T} > 130$ GeV.  The different treatment for the $e\mu$ channel is 
required due to the background composition of the $e\mu$ channel which includes diboson processes, and a Z+jets component 
dominated by $Z \rightarrow \tau\tau$ which has its own source of $E_{T}^{miss}$.

\section{Backgrounds}
Standard Model (SM) top anti-top production is the dominant and irreducible background as its final state is identical to that of a $t\bar{t}$ resonance.  
However, the additional non-top-pair SM background sources listed below can mimic the top pair final state and also contribute to the final selection:

\begin{itemize}
\item  Z+jets
\item  Wt production
\item  Diboson production
\item  "Fakes": QCD multijet and W+jets processes where one or two jets are misidentified as leptons.
\end{itemize}

The $Wt$ and diboson background predictions are determined from Monte Carlo. Fakes are estimated using a completely data-driven method \cite{ref:ttbarxsec}.  
The $Z+jets$ prediction derived from Monte Carlo is corrected for both its yield and jet multiplicity by normalizing to data within the control 
region (for $ee$ and $\mu\mu$ decays).  

\begin{figure}
\centering
\resizebox{1.0\columnwidth}{!}{%
  \centering
  \includegraphics{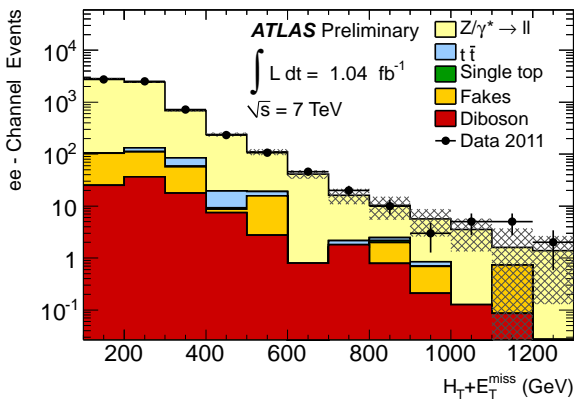}
 }
\caption{A comparison between Monte Carlo prediction and data for the $H_{T} + E_{T}^{miss}$ distribution in the control region for the $ee$ channel. 
The statistical uncertainty on the Monte Carlo prediction is represented by hatched bands\cite{ref2}.}
\label{fig:1}       
\end{figure}

We perform a validation of the $H_{T} + E_{T}^{miss}$ distribution in the data control sample as shown in 
figure \ref{fig:1}.  The expected background composition and the number of observed selected events in the signal region are shown in table \ref{table:bkg}. 
\begin{figure}
\centering
\resizebox{1.0\columnwidth}{!}{%
  \centering
  \includegraphics{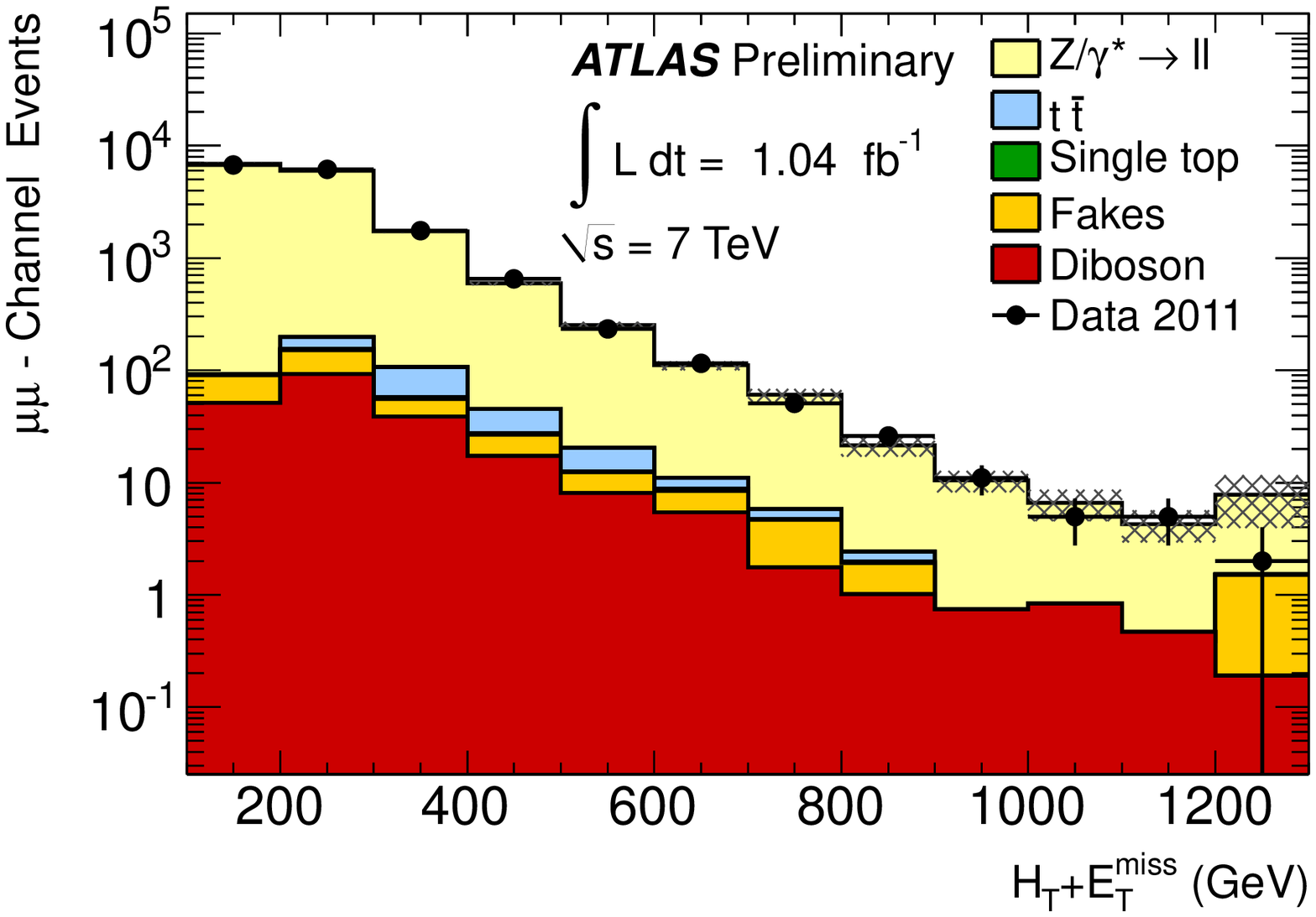}
 }
\caption{A comparison between Monte Carlo prediction and data for the $H_{T} + E_{T}^{miss}$ distribution in the control region for the $\mu\mu$ channel.
The statistical uncertainty on the Monte Carlo prediction is represented by hatched bands\cite{ref2}.}
\label{fig:2}       
\end{figure}

\begin{table}[h!]
    \centering
    \caption{Background composition in the signal region.
    Both statistical and systematic uncertainties are included \cite{ref:ttbarxsec}.}
    \begin{tabular}{|l|c|}
      \hline
      Process &  Predicted number of background events \\
      \hline
      $t\bar{t}$                  & 1920$^{+230}_{-220}$ \\
      $Z/\gamma^{*} \rightarrow ee$ + jets         & 130$^{+72}_{-49}$ \\
      $Z/\gamma^{*} \rightarrow \mu\mu$ + jets     & 140$^{+27}_{-21}$ \\          
      $Z/\gamma^{*} \rightarrow \tau\tau$ + jets   & 85$^{+12}_{-10}$ \\
      \rm{Diboson}                                      & 83$^{+13}_{-12}$\\
      \rm{Single top}                           & 98$^{+14}_{-13}$  \\
      \rm{Fakes}                                        & 96$^{+94}_{-51}$ \\
      \hline
      \hline
      Total background               &   2550$^{+330}_{-300}$  \\
      Data            &    2659 \\
      \hline
    \end{tabular}
    \label{table:bkg}
\end{table}


\section{Statistical Analysis}
\label{StatAnalysis}

\begin{figure}
\centering
\resizebox{1.0\columnwidth}{!}{%
  \centering
  \includegraphics{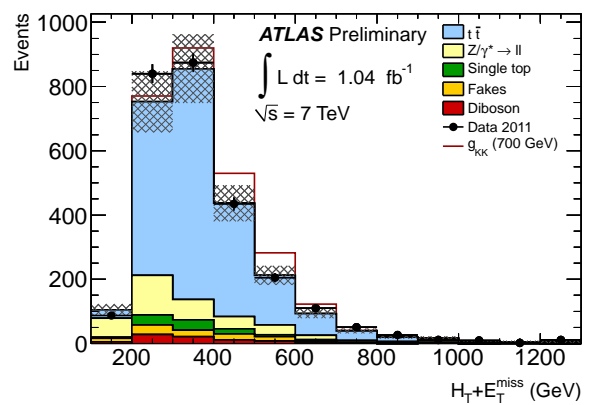}
 }
\caption{
Data -Monte Carlo comparison for the $H_{T} + E_{T}^{miss}$ distribution together with a KK-gluon signal
with a mass of 700 GeV for illustration. The statistical and systematic uncertainty on the Monte Carlo
is represented by the hashed band \cite{ref2}.}
\label{fig:3}       
\end{figure}

A deviation from the Standard Model is searched for using the discriminating variable $H_{T} + E_{T}^{miss}$. A comparison of the Monte Carlo prediction 
for the $H_{T} + E_{T}^{miss}$ distribution and observation in data, including a hypothetical signal, is shown in figure \ref{fig:3}. 

For the statistical analysis we construct a binned likelihood function 
for the number of signal and background events, based on Poisson statistics for the $H_{T} + E_{T}^{miss}$ 
distribution. 

\begin{eqnarray}
  {\cal L}(data|N_j,\theta_i) = \prod_{k=1}^{N_{bin}} \frac{\mu_k^{n_k} e^{-\mu_k}}{n_k!}\prod_{i=1}^{N_{sys}}G(\theta_i,0,1) \hspace{0.3cm} \nonumber \\ 
  where \hspace{0.3cm} \mu_k=\sum_j N_j T_{jk}(1+\theta_i \epsilon_{jik})
\label{eqn:lhoodtemp}
\end{eqnarray}

The likelihood depends on the expected number of events in each bin ($\mu_k$) which are the sum of the expected signal and background events.  
The indices $j=1$ and $j=2$ denote the signal and background templates respectively.  The dependence of the likelihood function on the hypothetical 
resonance mass is implicit via the template shapes ($T_{jk}$) which describe the fractional contribution of each bin to the overall yield. 
Thus, $\mu_k$ is determined by the expected total yields for signal and background ($N_j$), the background shape, and the resonance mass.  

Gaussian priors ($G$) are used for the systematics nuisance parameters ($\theta_i$) that control the bin-by-bin one 
sigma systematic variations ($\epsilon_{jik}$) that simultaneously alter the shape and yield of the templates.  Systematics are fully correlated between
signal and background.  Various sources of systematic uncertainty were studied to determine both rate and shape dependencies.  
A non-negligible impact on the analysis was found for the sources listed in Table \ref{table:syst}.  

\begin{table}[h!]
  \footnotesize  
  \centering
  \caption{Change in acceptance due to various sources of systematic uncertainties. Positive and negative acceptance variations are listed in [\%]. All signal systematic uncertainties have been symmetrized.
  The total systematic uncertainty for Standard Model background also includes luminosity (3.7\%) and the cross-section uncertainties \cite{ref2}.}
  \begin{tabular}{c|cc|c|}
    \cline{2-4}
                       & \multicolumn{2}{c|}{SM background} &  $m_{KK}$=700 GeV \\
                       & (+) & (-) & \\
    \cline{2-4}
    \hline
    \multicolumn{1}{|l|}{Lepton ID / Trigger}                   & 3.4 & 4.5 &  4.2 \\
    \hline
   \multicolumn{1}{|l|}{Jet energy scale}                       & 7.4 & 6.7 &  3.5 \\
    \hline
   \multicolumn{1}{|l|}{Jet energy resolution}                  & 2.3 & - &  2.5 \\
    \hline
    \multicolumn{1}{|l|}{ISR/FSR}                         & 0 & 2.3 & 2.5 \\
    \hline
     \multicolumn{1}{|l|}{Parton Shower}                & 1.4 & 1.4 & - \\
    \hline
     \multicolumn{1}{|l|}{Generator}                    & 4.8 & 4.8 &  -\\
    \hline
    \multicolumn{1}{|l|}{PDF}                           & 2.7 & 2.7 &  1.2 \\
    \hline
    \hline
    \multicolumn{1}{|l|}{Total Systematic}                           & 12.8 & 11.5 &  6.6 \\
    \hline
   \end{tabular}
   \label{table:syst}
\end{table}

A p-value quantifies the probability of observing an excess assuming the null hypothesis that is at least as great as what is observed in data. 
A p-value is evaluated using a log-likelihood ratio as a test statistic; a value of 44\% is obtained indicating that the data contains no 
statistically significant excess above the SM \cite{ref2}. 
 
\begin{figure}
\centering
\resizebox{1.0\columnwidth}{!}{%
  \centering
  \includegraphics{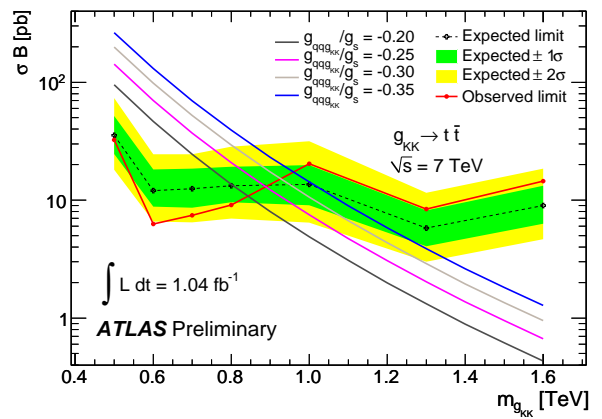}
 }
\caption{Expected and observed limits on cross section times branching ratio at 95
cross section for a Randall-Sundrum KK-gluon $g_{KK}$. Cross sections were calculated using the MRST
2007 LO$^{*}$ PDF \cite{ref2}.}
\label{fig:4}       
\end{figure}

Following the Bayesian prescription, a marginalized posterior probability density is obtained and used to determine 95\% C.L. upper 
limits on the number of signal events for a series of hypothesized resonance masses.  The limits on the number of signal events are converted into the 
limit on $\sigma B$ as a function of mass shown in figure \ref{fig:4}.
Corresponding limits on the KK-gluon ($g_{KK}$) mass are obtained and shown in 
table \ref{tab:massLimits} for the default RS1 model and a series of other scenarios where the coupling strength between the KK-gluon and the light 
quarks $(g_{qqg_{KK}})$ has been enhanced, thereby increasing the production cross section.  

\begin{table}[!htb]
  \centering
  \caption{Expected and observed lower limits on the KK-gluon mass in the Randall-Sundrum model \cite{ref2}.}
  \begin{tabular}{l|c|c|}
    \cline{2-3}
                        & \multicolumn{2}{|c|}{Mass Limit (TeV)}\\
    \hline
    \multicolumn{1}{|c|}{$g_{qqg_{KK}}/g_{s}$}  & Expected & Observed\\
    \hline
    \multicolumn{1}{|c|}{-0.20}         & 0.80 & 0.84  \\
    \multicolumn{1}{|c|}{-0.25}         & 0.88 & 0.88 \\
    \multicolumn{1}{|c|}{-0.30}         & 0.95 & 0.92 \\
    \multicolumn{1}{|c|}{-0.35}         & 1.02 & 0.96  \\
    \hline
  \end{tabular}
  \label{tab:massLimits}
\end{table}

\section{Results and Conclusion}
\label{ResultsAndConclusion}
A high mass $t\bar{t}$ resonance was searched for in 1.04 fb$^{-1}$ of data collected by the ATLAS detector.  The $H_{T}+E_{T}^{miss}$ 
distribution of the selected events in the signal region agrees well with the Standard Model prediction. No significant deviation is observed and   
limits on  $\sigma B$  are set for a series of resonance masses.

We exclude a KK-gluon resonance as predicted by the RS1 model with mass below 
840 GeV at 95\% C.L.  Stronger limits are set for cases where the coupling strength between the KK-gluon and light quarks is enhanced \cite{ref2}. 


\begin{thebibliography}{}
\bibitem{refTopSpec}
R. Frederix, F. Maltoni, JHEP \textbf{01}, (2009) 047.
\bibitem{ref1}
B. Lillie. et. al., JHEP \textbf{09}, (2007) 074.
\bibitem{ref2}
ATLAS Collaboration, ATLAS-CONF-2011-148. 
\bibitem{ref:atlasexp}
ATLAS Collaboration, JINST \textbf{3}, (2008) S08003.
\bibitem{ref:ttbarxsec}
ATLAS Collaboration, ATLAS-CONF-2011-117.
\end{thebibliography}
\end{document}